\documentclass[Journal,letterpaper,InsideFigs]{ascelike-new}
\usepackage[utf8]{inputenc}
\usepackage[T1]{fontenc}
\usepackage{lmodern}
\usepackage{graphicx}
\usepackage[style=base,figurename=Fig.,labelfont=bf,labelsep=period]{caption}
\usepackage{subcaption}
\usepackage{amsmath}
\usepackage{newtxtext,newtxmath}
\usepackage{bm}
\usepackage{float}
\usepackage{rotating, threeparttable, multirow, multicol, makecell, booktabs}
\usepackage{array}
\newcolumntype{M}[1]{>{\centering\arraybackslash}m{#1}}
\usepackage[colorlinks=true,citecolor=red,linkcolor=black]{hyperref}
\NameTag{Yang et al., \today}
\begin{document}
\nolinenumbers
\title{Machine learning-based multipoint optimization of fluidic injection parameters for improving nozzle thrust performance}

\author[1]{Yunjia Yang}
\author[1]{Jiazhe Li}
\author[1]{Yufei Zhang}
\author[1,*]{Haixin Chen}

\affil[1]{Tsinghua University, Beijing, 100084, People’s Republic of China.}

\affil[*]{Correspondence email: chenhaixin@tsinghua.edu.cn}

\maketitle

\begin{abstract}
Fluidic injection offers a promising solution to improve the performance of the overexpanded single expansion ramp nozzles (SERNs) during vehicle acceleration. However, determining the injection parameters that yield the best overall performance across multiple nozzle operating conditions remains a challenge. The gradient-based optimization method requires gradients of injection parameters at each design point, which can lead to high computational costs when using computational fluid dynamics (CFD) simulations. This paper uses a pretrained neural network to replace CFD during optimization, enabling quick calculation of the nozzle flow field at multiple design points. Considering the physical characteristics of the nozzle flow field, a prior-based prediction strategy is adopted to enhance the model’s accuracy. In addition, the neural network's back-propagation algorithm computes gradients quickly by running the computation only once, thereby greatly reducing gradient computation time compared to the finite difference method. As a test case, the average nozzle thrust coefficient of an SERN at seven design points is optimized, resulting in a 1.14\% improvement. The time cost is greatly reduced compared with traditional optimization methods, even when the time required to establish the training database is included.
\end{abstract}

\section{Practical Applications}

In aerodynamic design, early optimization studies often focused on a \textit{single operating condition}. However, real aircraft and propulsion systems must perform well across a wide range of conditions during acceleration, cruise, and off-design operation. \textit{Multi-point optimization} offers a more practical approach to optimizing overall performance but requires many expensive computational fluid dynamics (CFD) simulations. This study addresses this challenge by using a pretrained neural network as a fast surrogate for both performance and its gradients. We applied this method to design fluidic injection nozzles that perform well across the entire flight envelope.

\section{Introduction}

The single expansion ramp nozzle (SERN) is a critical component in wide-speed-range aerospace vehicles, generating the majority of engine thrust while significantly impacting vehicle lift and pitching moment \cite{lv_design_2019,yu_inverse_2021}. These vehicles operate at high altitudes and speeds, where ambient pressure is very low, resulting in a high nozzle pressure ratio (NPR) that often exceeds 100 \cite{snyder_design_1991}. To produce enough thrust, SERNs are designed with a large exit-to-throat area ratio to fully expand exhaust gases. However, during acceleration at low NPRs, this design can lead to overexpansion, reducing thrust performance and creating a thrust-minus-drag issue \cite{hirschel_design_2011,bowcutt_physics_2018} in transonic regions.

Over the past few decades, researchers have explored various methods to enhance SERN performance under overexpansion conditions, including movable flap \cite{gruhn_flap_2000,gruhn_improvement_2002}, contour optimization \cite{li_aerodynamic_2018,lv_design_2019}, passive cavity \cite{asbury_passive_1996}, energy deposition \cite{ju_optimization_2017}, and fluidic injection \cite{chow_interaction_1964,lv_numerical_2017,yang_analysis_2023}. Among these, fluidic injection stands out due to its efficiency, robustness, and seamless integration with the engine's secondary air system. Studies have examined the mechanisms and effects of injection parameters, including location, intensity, and angle. 

However, there remains a gap in practical methods for designing and optimizing these parameters. Given the wide-speed-range mission profile of the wide-speed-range vehicle, the nozzle will operate across a wide range of conditions, so the goal is to \textit{identify optimal injection parameters at multiple design points to maximize overall performance}. A major challenge in achieving this optimization is the high computational demand \cite{zhu_kriging-assisted_2024,yang_fast_2024}. Accurate performance evaluation relies on Reynolds-averaged Navier-Stokes (RANS) simulations, which become costly as the number of design points increases. Additionally, optimization requires calculating gradients of objective functions with respect to design variables, typically using finite difference (FD) or adjoint (AD) methods. The FD method requires at least one additional evaluation per design variable to compute the gradient. The AD method, although it requires only one CFD evaluation, has a computational cost for solving the adjoint problem comparable to that of CFD, and there is a possibility of divergence. 

Machine learning, especially deep neural networks, has emerged as a promising approach to overcoming the computational burden of multipoint optimization. One framework to apply machine learning models to optimization is data-based optimization (DBO) \cite{li_data-based_2019,yang_uncertainty-aware_2026}. It leverages pre-trained machine learning models trained on datasets encompassing typical geometries and conditions to rapidly predict performance without relying on repeated simulations. 

The DBO approach has been applied to optimization for airfoils \cite{renganathan_enhanced_2021,chen_numerical_2021,wang_airfoil_2024}. In this study, we extend it to optimize SERN injection parameters for improved overall performance across multiple flight conditions. Considering the properties of a nozzle flow field, we introduced several techniques to improve the model's effectiveness in the optimization process: 

\begin{enumerate}
    \item \textbf{High-dimensional outputs:} Instead of directly predicting performance metrics, we implement a U-Net model \cite{navab_u-net_2015} that outputs pressure and temperature distributions on the nozzle surfaces, from which performance metrics can be derived. This approach offers designers an intuitive insight during optimization (for instance, temperature distributions can inform cooling system design). The high-dimensional flow-field output also serves as a form of regularization, reducing model overfitting \cite{yang_fast_2024}.

    \item \textbf{Prior-based prediction:} Rather than directly predicting the target flow field from geometry and operating conditions, our model predicts the difference that the injection brings to the non-injection baseline flow field. This prior-based prediction strategy, introduced in our previous research \cite{yang_flowfield_2022}, is effective since the influence of injection remains consistent across various nozzle conditions and parameters, and additionally, most of the flow field remains unchanged with or without injection.

    \item \textbf{Gradients via backpropagation:} While many DBO  use pretrained models solely for forward prediction, one advantage of neural networks is that it is also possible to quickly obtain the gradients via backpropagation (BP) \cite{chen_numerical_2021}. These gradients can also be used in optimization and greatly reduce gradient calculation time, which is the major challenge in multipoint nozzle injection optimization.
\end{enumerate}

In the following sections, the mechanism of fluidic injection in SERNs is first introduced to define the engineering background, followed by a mathematical definition of the proposed optimization problem. Then, the dataset and the prior-based U-Net model designed for this problem are elaborated, and their performance is demonstrated. Finally, a gradient-based optimization with backpropagation is introduced, and its performance on a 7-design-point problem is analyzed.

\section{Mechanism of fluidic injection to SERN performance}

\begin{figure}[ht]
    \centering
    \includegraphics[width=0.5\linewidth]{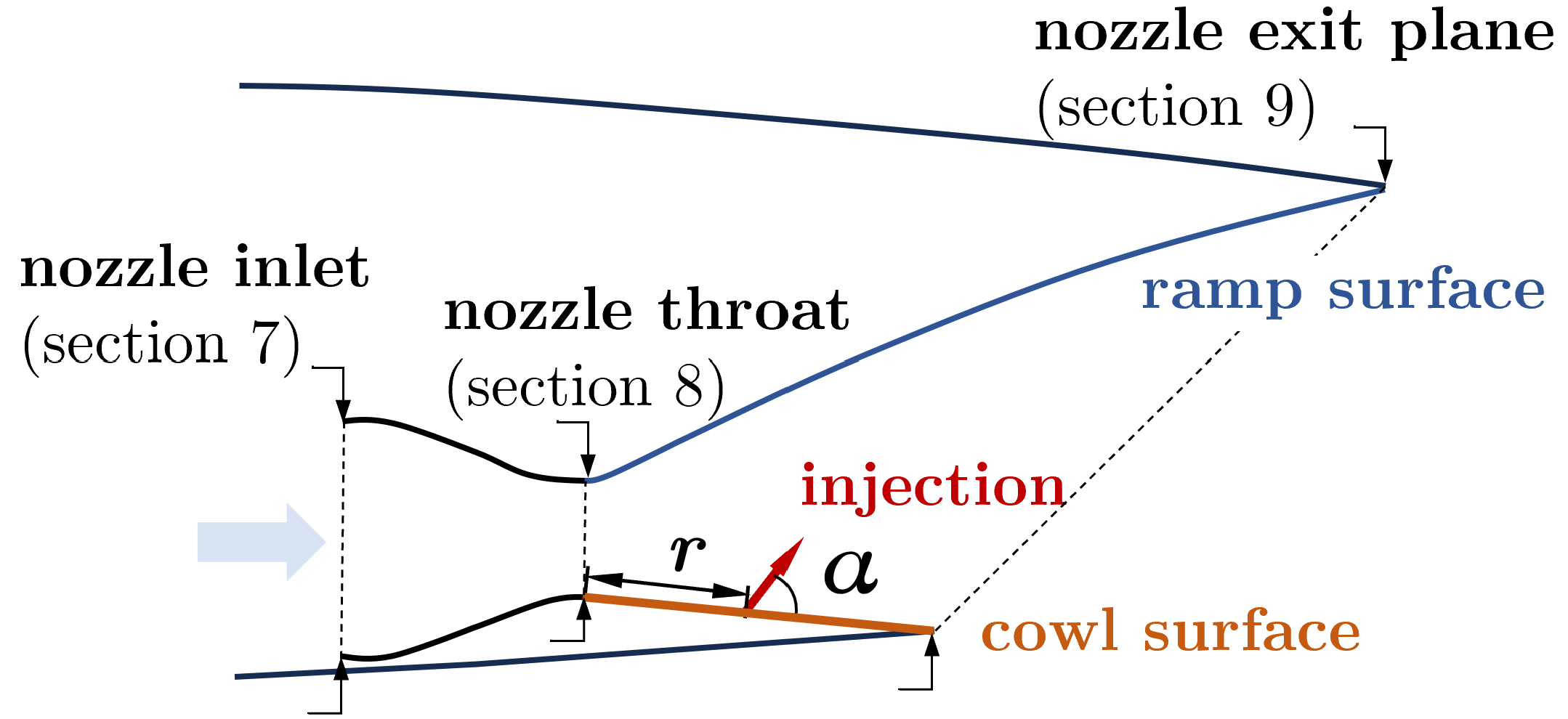}
    \caption{Visualization of nozzle geometry, the injection location, and angle}
    \label{fig:shapes}
\end{figure}

A two-dimensional SERN, as shown in Fig. \ref{fig:shapes}, is a key thrust component for a wide-speed-range vehicle. There are three control sections in the nozzle flow passage: the inlet (Section 7), the throat (Section 8), and the exit plane (Section 9). The segment between the inlet and throat is the convergent segment, where the flow is subsonic, and the segment afterward is the divergent segment, where the flow accelerates to supersonic. The upper side of the divergent segment (ramp) is much longer than the lower side (cowl) to enable sufficient expansion while saving weight. 

\begin{figure}[ht]
\centering

\begin{subfigure}{0.49\linewidth}
\centering
\includegraphics[width=\linewidth]{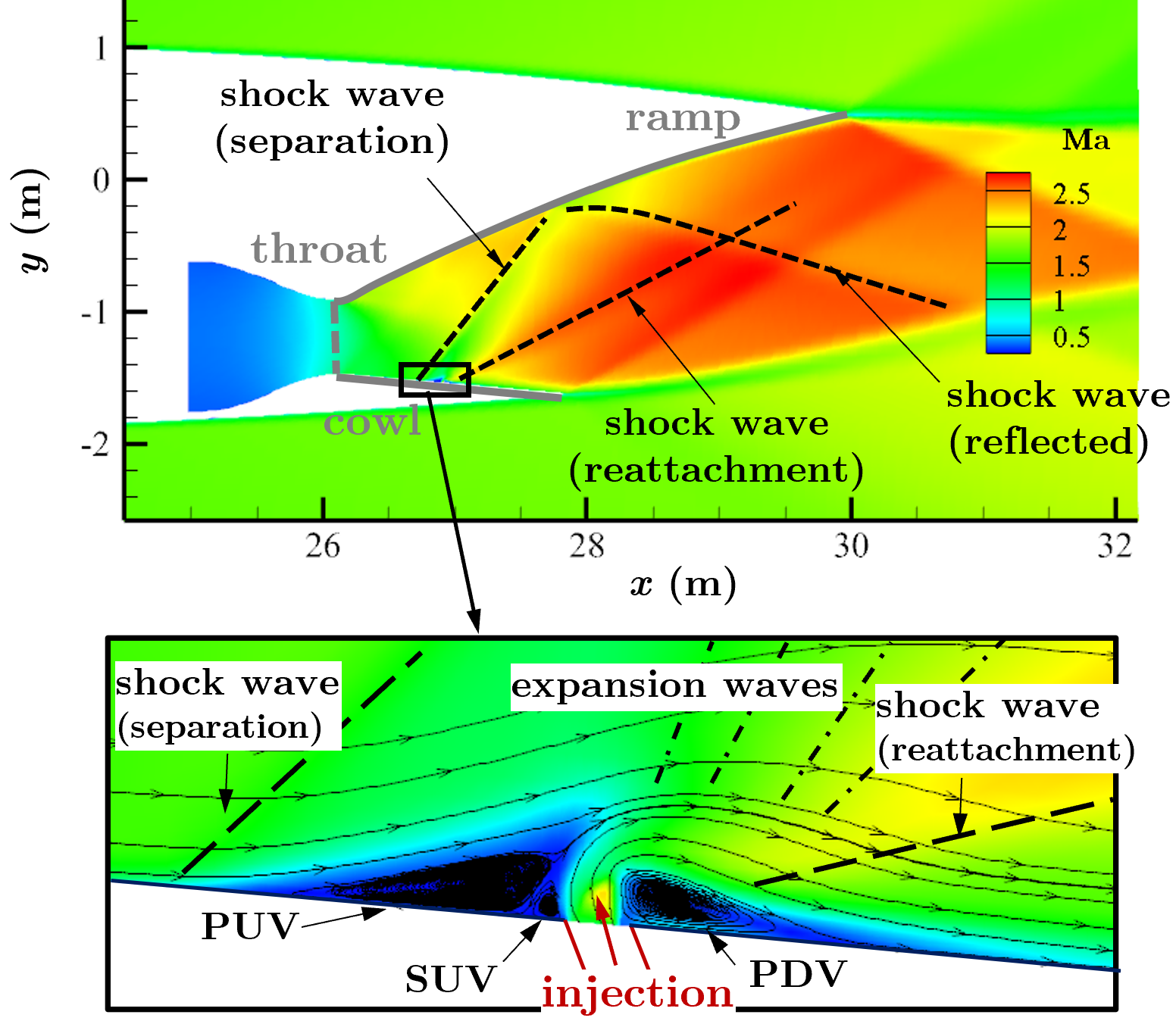}
\end{subfigure}
\begin{subfigure}{0.49\linewidth}
\centering
\includegraphics[width=\linewidth]{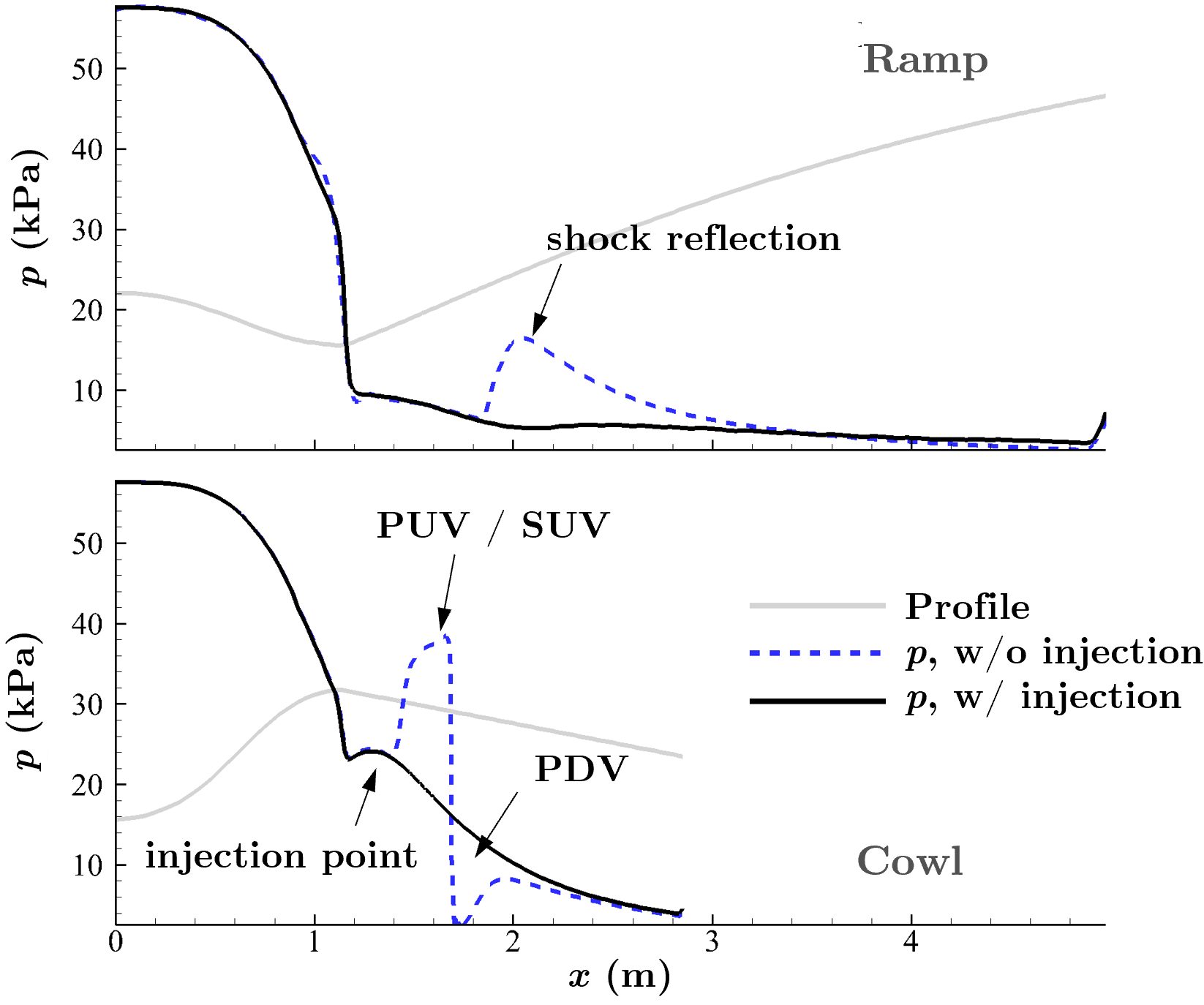}
\end{subfigure}

\caption{Mechanism sketch of a SERN with fluidic injection: (\textbf{left}) Flow field with fluidic injection. (\textbf{right}) Pressure distributions on ramp and cowl with and without injection.}\label{fig:mech}
\end{figure}

The single injection on the cowl surface is considered in this study following previous studies \cite{lv_numerical_2017,shanmugaraj_effects_2020,yang_analysis_2023} for its good efficiency in improving the nozzle’s thrust. Figure \ref{fig:mech} depicts an overexpanded SERN flow field with a typical fluidic injection and the pressure distributions comparison on the nozzle surfaces with and without fluidic injection. The results are obtained by the numerical method introduced in the following section. 

It can be observed that the penetrating injection forms a barrier to the mainstream, and a separation zone forms upstream. It consists of two vortices due to shear stress: the primary upstream vortex (PUV), which rotates clockwise, and the secondary upstream vortex (SUV), which rotates counterclockwise. The upstream vortex is similar to a wedge inserted into the mainstream, producing an oblique shock wave ahead of the separation point. The shock wave deflects the mainstream and increases its pressure, forming a high-pressure zone on the cowl surface. Moreover, the shock wave propagates toward the ramp surface, resulting in a pressure peak.

The injection plume is then forced to deflect and adhere to the wall, and during this process, expansion waves emerge in the mainstream. They gradually reduce the pressure peak caused by the shock wave when it hits the ramp. On the cowl surface, a clockwise rotating primary downstream vortex (PDV) is formed behind the slot, where the pressure is lower than the baseline pressure. At the end of the separation, the mainstream reattaches, and another oblique shock wave forms, turning the mainstream and restoring pressure.

The effect of fluidic injection on nozzle performance can be divided into two aspects: the reaction force generated by the injection itself and the force resulting from integrating the injection's influence on surface pressure distributions. 

Quantitatively, the nozzle thrust equals the sum of the momentum flux through the throat and the integral of pressure on the expansion wall surfaces as
\begin{align}
    \bm{F} & =\int_{A_8} \rho \bm{V} (\bm{V} \cdot \bm {n}) + (p - p_{\mathrm{atm}}) \bm {n} \cdot \mathrm{d}\Gamma+ \int_{\mathrm{wall}}(p-p_{\mathrm{atm}}) \bm {n} \cdot \mathrm{d}\Gamma \nonumber \\ &= \bm {\Phi}_8 + \int_{\mathrm{wall}} (p-p_{\mathrm{atm}}) \bm {n} \cdot \mathrm{d}\Gamma,\label{eqn:f}
\end{align}
where $\bm {\Phi}_8$ is the momentum flux through the nozzle throat. $A_8$ is the throat area and, in two-dimensional circumstances, is the height of the throat area. Suppose that the new pressure distribution after introducing fluidic injection is $p'$. The new thrust will be:
\begin{equation}\label{eqn:injf}
    \bm{F}'= \bm {\Phi}_8' + \int_{\mathrm{wall}} (p' - p_{\mathrm{atm}}) \bm n \cdot \mathrm{d}\Gamma + \bm {\Phi}_s,
\end{equation}
where $\bm {\Phi}_s$ is the momentum flux through the injection slot. Since the injection will not influence upstream in a supersonic flow field, we have $\bm {\Phi}_{8} \approx \bm {\Phi}_{8}'$.

The $x$-direction thrust coefficient $C_{f,x}$ is used as an indicator of nozzle performance. Its definition under non-injection conditions is: $C_{f,x} = F_x / F_{\mathrm{id}}$
where the ideal thrust $F_{\mathrm{id}}$ is
\begin{equation}\label{eqn:ideal}
    F_{\mathrm{id}} = \dot m_8 \sqrt{\frac{2\gamma R T_7^*}{\gamma - 1}\left(1 - \left(\frac{p_7^*}{p_{\mathrm{atm}}}\right)^{-\frac{\gamma - 1}{\gamma}}\right)},
\end{equation}
where $\dot m_8$ is the mass flow rate through the throat and $p_7^*$ and $T_7^*$ are the total pressure and temperature at the nozzle inlet, respectively. In the present paper, the ideal gas with $\gamma=1.4$ is used for calculating the ideal thrust for simplicity, whereas all CFD simulations for generating the dataset use a variable $\gamma$. However, the resulting discrepancy applied consistently to all cases. The relative trends are therefore still considered meaningful, while the thrust coefficient value should be interpreted as a normalized performance indicator rather than an exact thermodynamic efficiency. 

For a nozzle with fluidic injection, the ideal thrust should include the injection contribution, since it is always supplied by the compressor or inlet bleeding and can expand in the nozzle. Then, the thrust coefficient is written as $C_{f,x}' = F_{x}' / (F_{\mathrm{id}} + F_{\mathrm{id}}')$,
where $F_{\mathrm{id}}'$ is the ideal thrust of fluidic injection, which is calculated similarly to Eq. \ref{eqn:ideal} with the total condition of the injection. 

From Eqs. \ref{eqn:f} and \ref{eqn:injf}, if
\begin{equation}
    F_{\mathrm{id}}' < \Delta F_x = \int_{\mathrm{wall}} (p' - p) \bm n \cdot \mathrm{d}\Gamma + \bm {\Phi}_s,
\end{equation}
the fluidic injection will improve the nozzle's performance.

\section{Problem setup}

The aforementioned analysis demonstrated that proper fluidic injection improves nozzle performance. However, as nozzle operating conditions change with the vehicle's flight conditions, the need for fluidic injection to best enhance the nozzle’s performance also varies. It remains a design challenge to determine optimal injection parameters across the entire flight profile, especially given that the nozzle operating conditions vary dramatically.

Specifically, we fix the nozzle contour $g$ and select a series of design points $\mathcal C = \left\{c_i\right\}_{i=1, \cdots, N_c}$ for the desired flight profile, and the design variables includes fixed ones ($\theta$) and ones that can be varied with operating condition ($\theta(c)$). The optimization problem is formulated to find the best parameters that achieve the largest average thrust coefficient as
\begin{equation}
    \max_{\theta, \theta(c)} \sum_{c \in \mathcal C} C_{f}'\left(g,c\right).
\end{equation}

Below, we provide details on the settings.

\begin{itemize}
    \item \textbf{SERN profile ($g$)}: The fixed SERN contour shown in Fig. \ref{fig:shapes} is selected as the baseline configuration $g$ in the present paper. The profile was generated with \texttt{AeroMOC} (\href{https://github.com/YangYunjia/AeroMOC}{https://github.com/YangYunjia/AeroMOC}). It utilizes the asymmetric two-dimensional maximum thrust method based on the method of characteristics \cite{mo_research_2015}. Boundary-layer correction and linear truncation are then applied to produce the final design. Specifically, the SERN is 2D, and the heights of the inlet, throat, and exit plane are 1.02 m, 0.53 m, and 2.12 m, respectively. The lengths of the two surfaces in the $x$ direction from the throat are 1.66 m and 3.83 m, respectively.
    
    \item \textbf{Nozzle operating conditions ($c$)}: During vehicle acceleration, the flight Mach number ($Ma$) and altitude ($H$) vary and determine the nozzle outlet condition. The nozzle’s inlet condition is described with the pressure ratio ($NPR$), defined with total inlet pressure and ambient pressure $NPR = p_7^*/p_{\mathrm{atm}}$, and total inlet temperature ($T_7^*$). 
    
    \item \textbf{Fluidic injection parameters ($\theta$)}: Four parameters determine the injection conditions: injection location ($r$), injection angle ($\alpha$), and total inlet temperature of the injection ($T_s^*$) are fixed across design points, while the secondary pressure ratio ($SPR$) can vary using the adjusting valves in the secondary air system. Specifically, $r$ is defined as the proportional station on the cowl flap surface, and $\alpha$ is defined as the angle between the injection and the tangent of the cowl flap surface. An illustration can be found in Fig. \ref{fig:shapes}. The slit width was fixed at 2 cm. $SPR$ is defined as the ratio between the injection total pressure and mainstream inlet total pressure $SPR = p_s^*/ p_7^*$.
    
\end{itemize}


\section{Database establishment}

A database is established to train the machine learning model. The database contains flow fields under different nozzle conditions and injection parameters, enabling the trained model to be applied to optimization tasks.

\subsection{Sampling of the nozzle conditions and injection parameters}

The Latin hypercube sampling (LHS) method is used to generate 300 nozzle operating conditions. The ranges of the operating conditions are determined by the common flight envelope of the wide-speed-range vehicle and are listed in Table \ref{tab:ranges}. Since $NPR$ is largely determined by the vehicle speed, its sampling range is set based on $Ma$ for each sample. 

For each nozzle operating condition, 36 injection conditions are obtained via uniform sampling to improve extendability. Table \ref{tab:ranges} also shows the ranges of the parameters. Note that the injection temperature range is related to the mainstream inlet total temperature.

\begin{table}[ht] 
\small 
\caption{Ranges of nozzle operating conditions and injection parameters.}\label{tab:ranges}
\centering
\begin{tabular}{ccM{4cm}M{4cm}}
\toprule
& & \textbf{lower boundary} & \textbf{upper boundary} \\
\midrule
\multirow{4}{*}{\makecell{nozzle\\ operating\\ conditions}} & $Ma$ & 1.5 & 3.5 \\
& $H$ (km)& 15& 20\\
& $NPR$ & $3 + 12 \times (Ma-1.5)/2$ & $10 + 20 \times (Ma-1.5)/2$ \\
& $T_7^*$  (K)& 900& 2000\\
\midrule
\multirow{4}{*}{\makecell{injection\\ parameters}}& $r$& 0.10& 0.95\\
& $\alpha$ (°)& 30& 150\\
& $SPR$& 0.100& 2.375\\
& $T_s^*$ (K)& 300&  $300+0.6\times(T_7^*-300)$\\
\bottomrule
\end{tabular}
\end{table}

In total, 10,800 parameter sets are obtained. We also simulated the non-injection flow fields under 300 nozzle operating conditions.

\subsection{CFD simulations and post-processing}

The two-dimensional, steady, and compressible Reynolds-averaged Navier–Stokes (RANS) equations are solved with the finite volume method. A total-variation-diminishing framework based on multi-dimensional interpolation and a modified Riemann solver is used for spatial discretization, and an implicit scheme is used for time integration. The realizable k-$\epsilon$ model is commonly used for supersonic flows, especially for secondary injection into a supersonic crossflow \cite{huang_influences_2012}. The real gas is used for the simulations in the present study, and the temperature-dependent specific heat capacity was evaluated using a piecewise NASA polynomial formulation. 

The simulations are conducted on a structured grid. The inlet boundaries of the nozzle are specified by the given nozzle operating conditions, and the external-flow boundaries are determined by the characteristics of the Riemann invariants. The no-slip adiabatic condition is imposed on all the wall surfaces. The secondary flow passages are not included in the computational region, and injection is applied using a total pressure–total temperature boundary condition at the injection slot's exit plane. The injection angle is set by imposing the velocity direction of the boundary condition. This method can reduce the grid complexity and has been used in a similar study \cite{erdem_numerical_2010,wang_prediction_2017}.

\begin{figure}[ht]
\centering
\begin{subfigure}{0.35\textwidth}
    \centering
    \includegraphics[width=\linewidth]{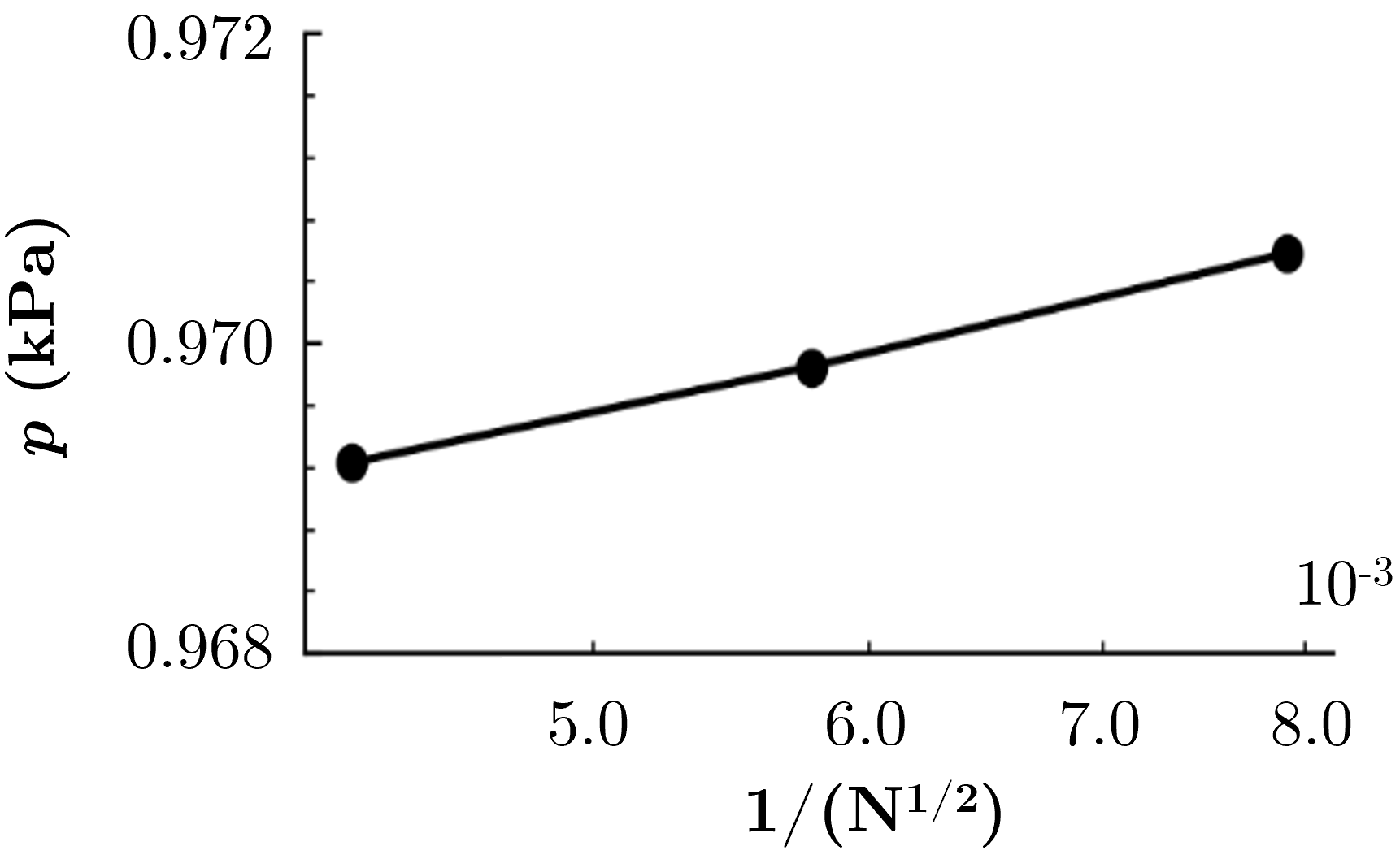}
\end{subfigure}
\begin{subfigure}{0.55\textwidth}
    \centering
    \includegraphics[width=\linewidth]{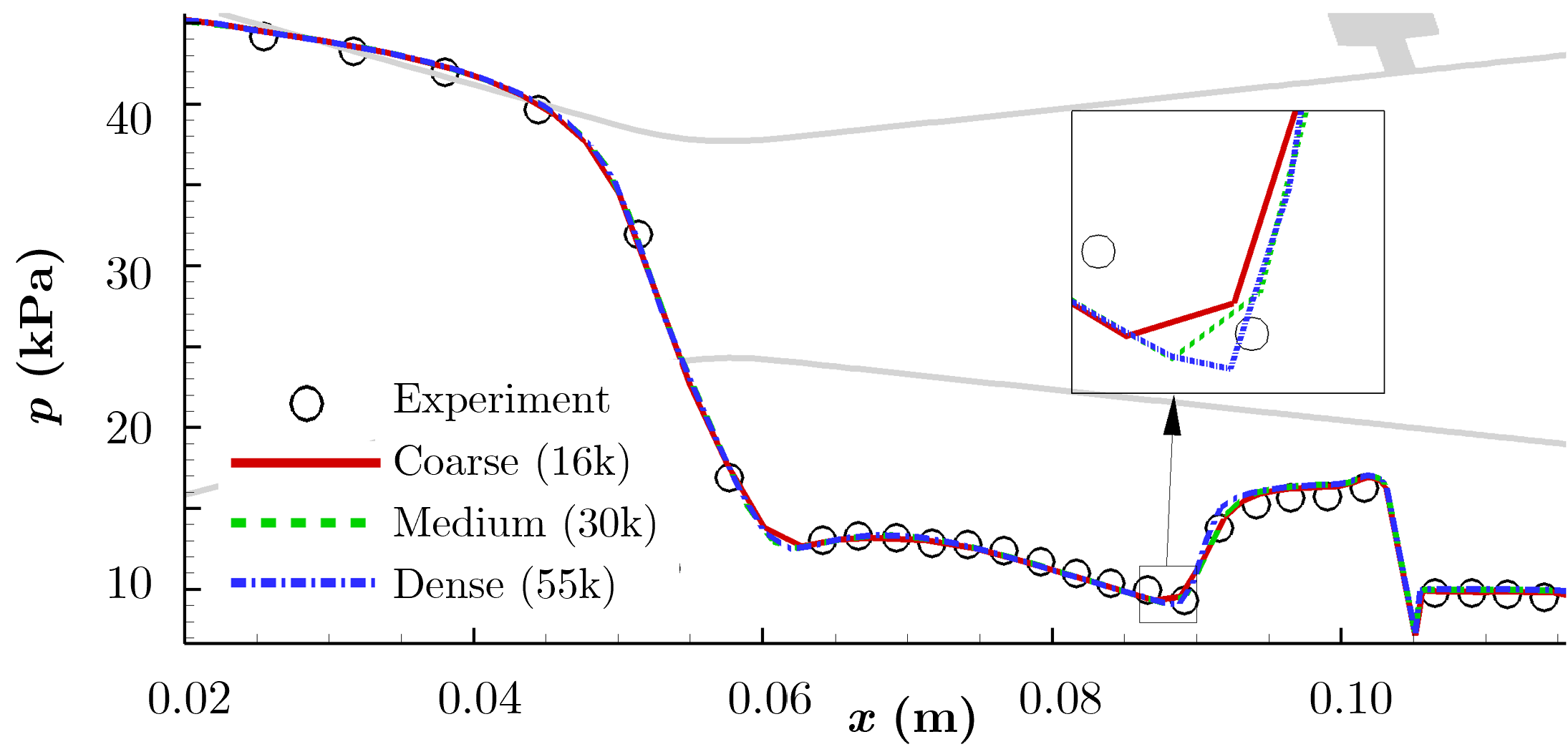}
\end{subfigure}

\caption{Grid convergence study for three different grids for the fluidic thrust vectoring two-dimensional nozzle: (\textbf{left}) Pressure distributions on the upper surface. (\textbf{right}) Thrust coefficient.} \label{fig:grid}
\end{figure}

The simulation methodology is validated using experimental data for a two-dimensional fluidic thrust-vectoring nozzle contributed by \citeNP{waithe_experimental_2003}. In their study, secondary injection was introduced in the divergent section on the upper nozzle surface, inducing oblique shock waves and flow separation. 
The simulation was conducted with three grid sizes. The wall pressure distributions on the upper surface for different grids are depicted in Fig. \ref{fig:grid}, together with the experimental results obtained in \citeNP{waithe_experimental_2003}. A quantitative study is also presented in which the nozzle thrust coefficients $C_f$ are integrated from the pressure distributions and plotted against the equivalent grid size $1/(N^{1/2})$, where $N$ is the number of grid points. It indicates that different grids lead to negligible differences in key flow features. The medium grid is selected to balance the accuracy and computation cost.

\begin{figure}[ht]
    \centering
    \includegraphics[width=0.9\linewidth]{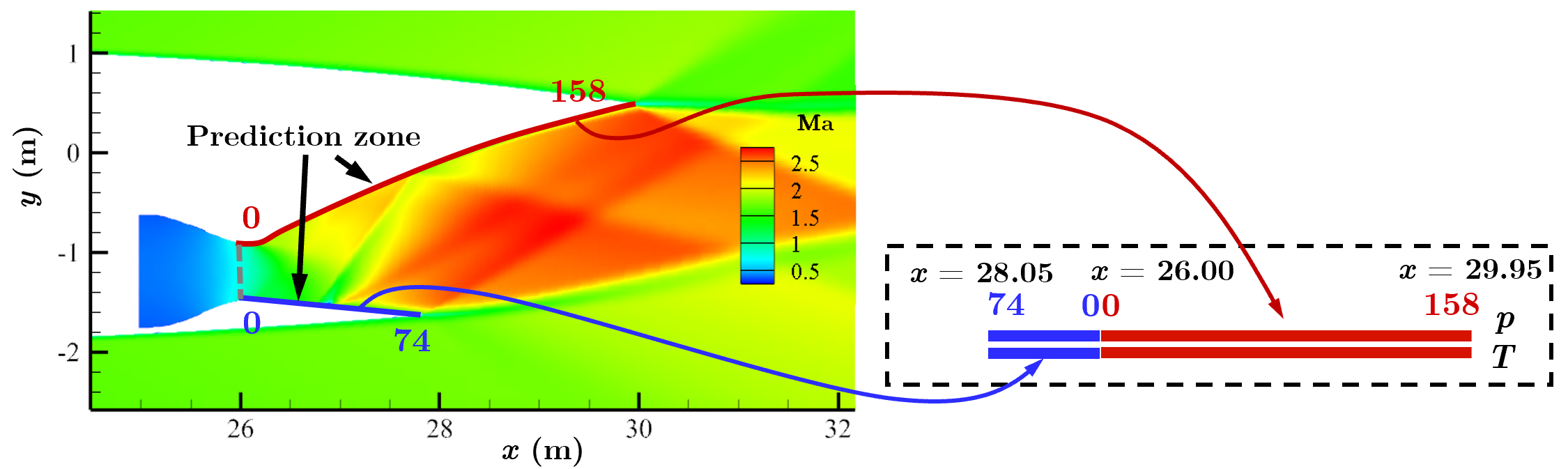}
    \caption{Visualization of data postprocessing.}
    \label{fig:probe}
\end{figure}

The pressure and temperature distributions on both the nozzle ramp and the cowl surfaces are used as the model's primary output. With the additional data of momentum flux ($\bm \Phi_8$), and nozzle contour, the thrust coefficient can be reconstructed as detailed in Appendix \ref{app:coef}. To unify the different meshes across the samples, we interpolate the raw solver output onto a series of probe points on the surfaces, equidistant in the $x$ direction. These points are then linked at the throat as visualized in Fig. \ref{fig:probe} to form an array of $2 \times 234$ for model prediction.

In addition, both the injection parameters and the distributions are normalized to improve model training. For the parameters, this is done with the upper and lower boundaries, whereas for the distributions, the total conditions at the nozzle inlet are used. The nondimensional values are

\begin{equation}
    \tilde{p} = \frac{p_7^* - p}{p_7^* - p_{\mathrm{atm}}} = 1 - \frac{p / p_{\mathrm{atm}} - 1}{NPR - 1}, \quad \tilde{T} = \frac{T_7^* - T}{T_7^* - T_s^*} \approx \eta,
\end{equation}
where $\eta$ is the adiabatic cooling efficiency.

\section{Pre-trained machine-learning model}

The prior-UNet (pUNet) model proposed in our previous study \cite{yang_fast_2024} is used to predict the nozzle surface flow field. It leverages a prior-based prediction strategy and the UNet framework to improve its generalization. 

\subsection{Prior-based prediction strategy}

\begin{figure}[ht]
\centering
\subfloat[\centering]{\includegraphics[width=0.6\textwidth]{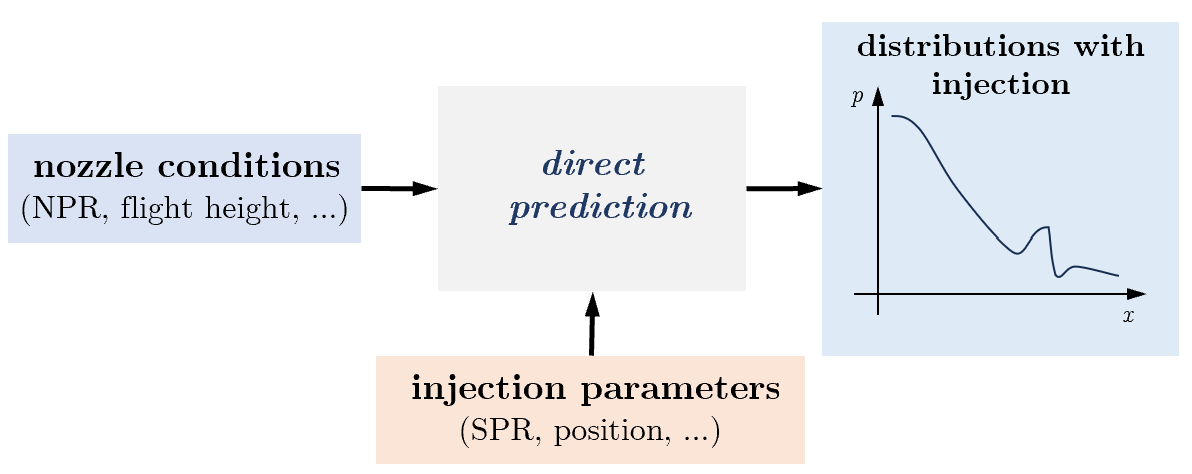}}\\
\subfloat[\centering]{\includegraphics[width=\textwidth]{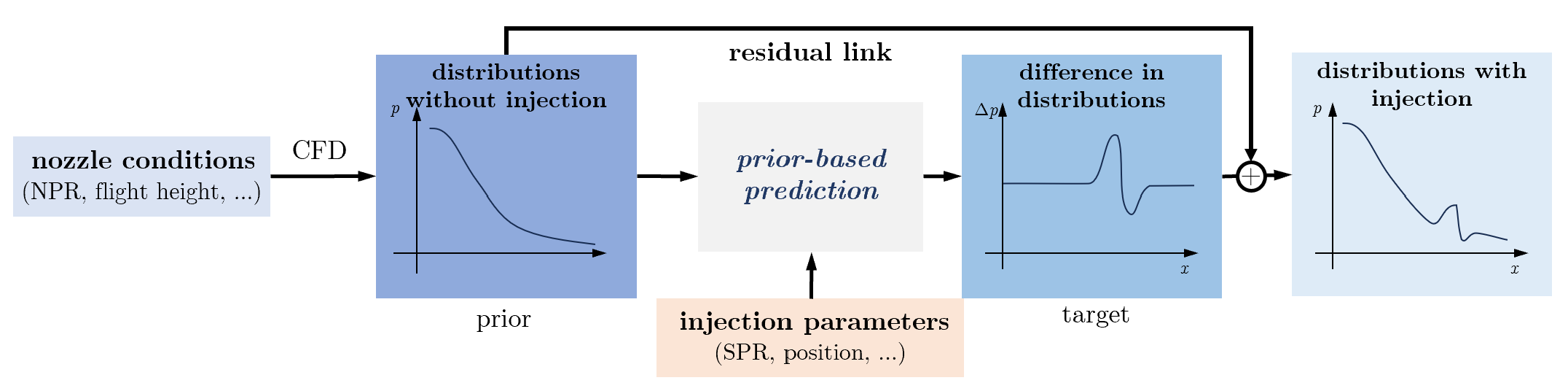}}
\caption{Model frameworks: (\textbf{a}) Direct prediction. (\textbf{b}) Prior-based prediction.}\label{fig:frame}
\end{figure}

Figure \ref{fig:frame} (a) shows the most straightforward approach for predicting the flow field. It utilizes a decoder-only model that takes the primary input (i.e., nozzle operating condition and injection parameters) and output (surface distributions). Prior-based prediction strategy introducing a \textit{reference flow field} as a prior and predicting the residual part between the target and the reference flow field. The reference flow field can be a low-fidelity version of the target flow field \cite{wu_generative_2022,zhang_multi-fidelity_2023} and can also be a flow field of the same geometry but under different operating conditions \cite{yang_flowfield_2022}. The common principle is that \textit{the prior and target flow fields share a strong, constant relationship}, which simplifies the mapping the model is learning.

In the present work, the surface distributions without injection (non-injection flow field) are used as reference flow fields when predicting the surface distributions after a given fluidic injection is introduced. The physics behind this choice lies in the supersonic characteristics of the flow field, which ensure that most of the injection flow field remains the same as the non-injection flow field. The prior-based approach to predicting the distributions is illustrated in Fig. \ref{fig:frame} (b). For each nozzle operating condition, a CFD simulation is first performed to obtain the surface pressure and temperature distributions without fluidic injection. Then, the residual model predicts the distributions after any parameters are injected into the flow field.

\subsection{Convolutional U-Net architecture}

\begin{figure}[ht]
    \centering
    \includegraphics[width=1\linewidth]{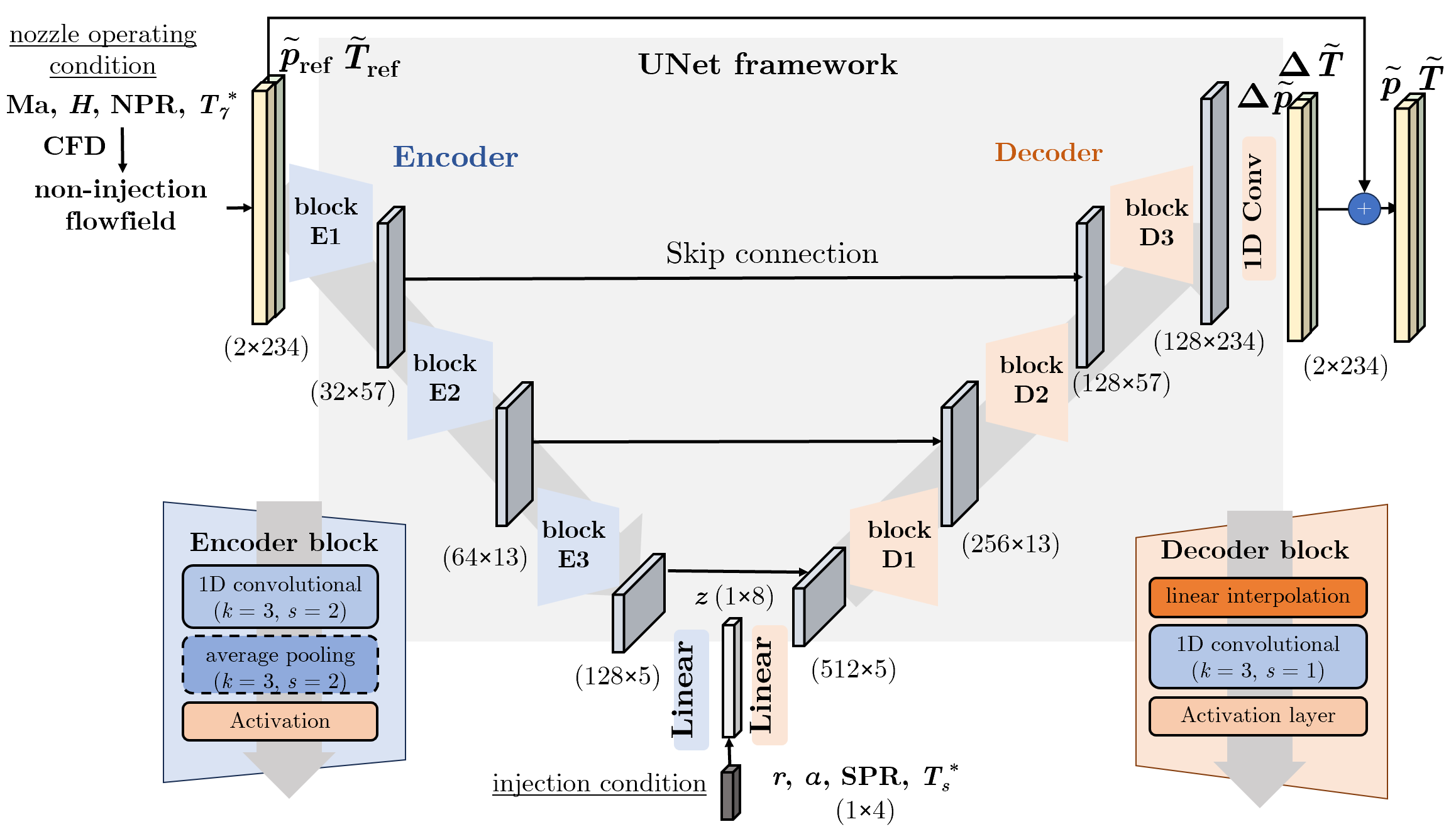}
    \caption{Convolutional U-Net architecture for nozzle surface flow prediction.}
    \label{fig:unet}
\end{figure}

The prior-based prediction is implemented with a convolutional U-Net model as depicted in Fig. \ref{fig:unet}. The encoder accepts the noninjection pressure and temperature distributions as inputs and extracts a low-dimensional representation $\bm z$. The decoder then combines $\bm z$ with the injection parameters and generates the residual in the two distributions. The feature maps along the contracting path are saved and reinjected into the upscaling path by concatenating the information from the symmetric level. 

One-dimensional convolutional layers are used to construct both the encoder and the decoder. Both encoder and decoder comprise three blocks. Each encoder block contains a 1D convolution layer with kernel size 3 and stride 2, an average pooling layer with the same kernel size and stride, a batch normalization layer, and an activation. A densely connected layer then links the flattened output to $\bm z$. The decoder block similarly consists of a linear interpolation layer for upsampling the 1D feature map, a convolution layer with a stride of 1, and batch normalization and activation. The network concludes with another convolution layer with a stride of 1 to compress the last feature map to two channels. The Gaussian error linear unit (GELU) is used as the activation function. 

The model hyperparameters are determined through several trial-and-error tests to achieve a balance between model performance and the size of trainable parameters. The channel numbers after each encoder block are 32, 64, and 128. For the feature maps after each decoder block, the channel counts are 512, 256, 128, and 128, respectively. The latent dimension before concatenating with the injection conditions is 8. This results in a total of 611,498 trainable parameters. 


\subsection{Model training}

The data are divided into a training set containing samples from 270 nozzle operating conditions, with the remaining samples held out for testing. The mean square error (MSE) between the model-predicted and CFD-simulated pressure and temperature distributions is used as the loss function. For optimization, a fixed batch size of 16 and the Adam algorithm are used. The warmup strategy increases the learning rate from $1 \times 10^{-5}$ to $1 \times 10^{-4}$ over the first 20 epochs to avoid instability at the beginning of training. The learning rate is subsequently reduced by an exponential function with a base of 0.95.

The training process is run three times to cross-validate the model. In each run, 10\% of the training samples are randomly selected for validation, and each run starts with a randomly initialized model. During training, the losses on the training and validation sets are monitored to avoid overfitting, and all three runs converge after 300 epochs.

\subsection{Model performance}

We compare the proposed pUNet model with a decoder-only direct prediction model with the same decoder implementation and training settings. The mean absolute errors (MAEs) of $\tilde p$ and $\tilde T$ are obtained by averaging across samples and data points 
and are reported in Table \ref{tab:errors}. We also reconstruct the ground-truth and predicted thrust coefficients ($C_{f,x}$) from $\tilde p$. The proposed pUNet model reduces errors by 36.1\% and 11.9\% in the pressure and temperature distributions on the test dataset. The improvement is more obvious in the averaged prediction accuracy of $C_{f,x}$, where the pUNet model reduced the error by 54.3\%. The pUNet model especially outperforms on the test dataset, further demonstrating the benefits of introducing a prior reference field for the model's generalization. Considering all three quantities have values around one, the final prediction errors are low enough for optimization. 

\begin{table}[ht] 
\small %
\caption{Prediction performance of the proposed machine learning model.}\label{tab:errors}
\centering
\begin{tabular}{ccccc}
\toprule
& & $\bm{\Delta \tilde p}$ & $\bm{\Delta \tilde T}$ & $\bm{\Delta C_{f,x}}$\\
\midrule
\multirow{2}{*}{Decoder-only} & training set & $0.00338 \pm 0.00002$ & $0.00434 \pm 0.00004$ & $0.00168 \pm 0.00022$\\
 & testing set & $0.00360 \pm 0.00002$ & $0.00445 \pm 0.00004$ & $0.00197 \pm 0.00023$\\
\multirow{2}{*}{pUNet} & training set & $0.00229 \pm 0.00008$ & $0.00395 \pm 0.00008$ & $0.00089 \pm 0.00004$\\
 & testing set & $0.00230 \pm 0.00009$ & $0.00392 \pm 0.00009$ & $0.00090 \pm 0.00006$\\
\bottomrule
\end{tabular}
\end{table}

\begin{figure}[ht]
    \centering
    \includegraphics[width=18cm]{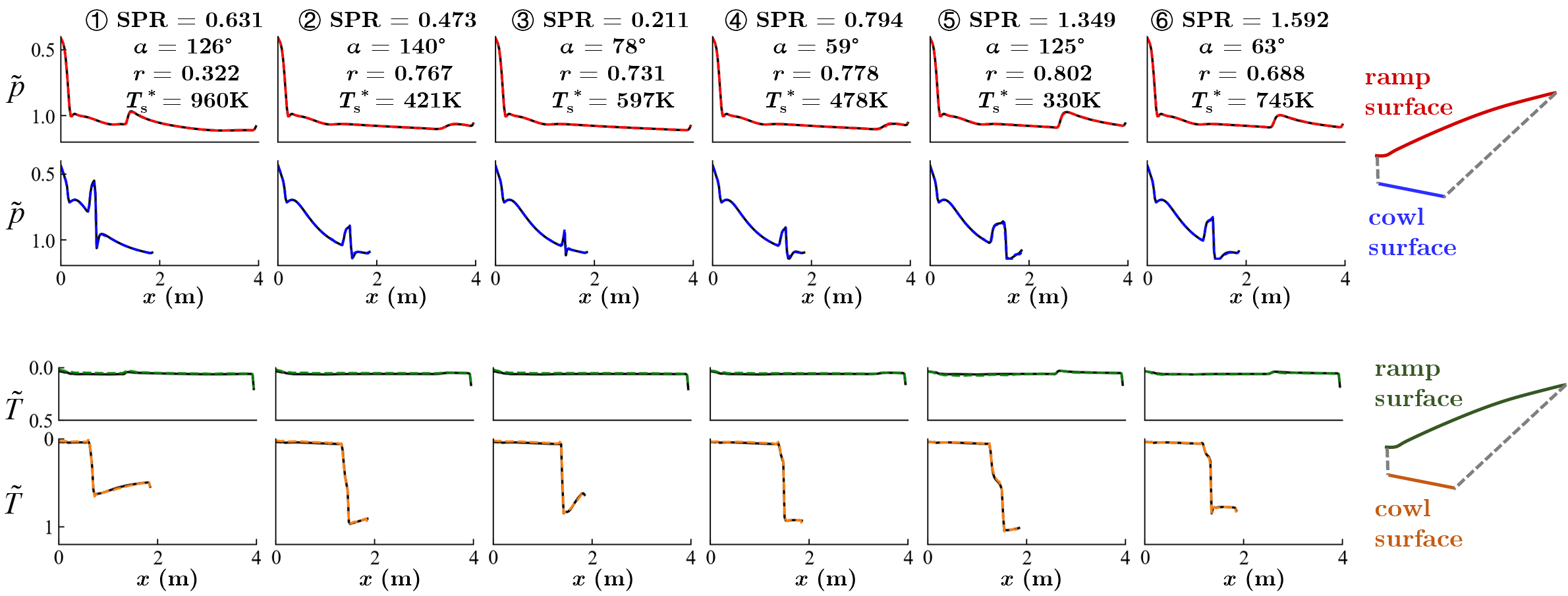}
\caption{Visualization of model-predicted surface pressure and temperature distributions. (from top to bottom in each column shows the $\tilde p$ on the ramp and cowl, the $\tilde T$ on the ramp and cowl successively. The model predictions are shown in colored dashed lines, and the ground-truth values are shown in black solid lines)}\label{fig:contours}
\end{figure}

We also provide the model-predicted distributions of nozzle surface pressure and temperature in Fig. \ref{fig:contours}. The nozzle operating conditions are randomly selected from the test samples: $Ma = 2.48$, $H = 17.5$ km, $NPR = 12.4$, and $T_7^* = 1917.6$ K, and the injection parameters are also randomly selected, which are shown at the top of each column. The model predictions are plotted as colored dashed lines, and the ground truth values are shown as black solid lines. The results further demonstrate that the pre-trained model is practical for optimization.

\section{Model-based multipoint optimization}

This section uses the pre-trained model for a practical multipoint optimization based on the flight profile.

\subsection{Gradient-based optimization via back-propagation of neural network}

One key advantage of a neural network surrogate model is that the gradients of the optimization objectives with respect to the design parameters can be easily computed via the BP algorithm, just as the tunable weights and biases in the neural network model are optimized during model training.

\subsubsection{Gradient via back-propagation}

\begin{figure}[ht]
    \centering
    \includegraphics[width=0.7\linewidth]{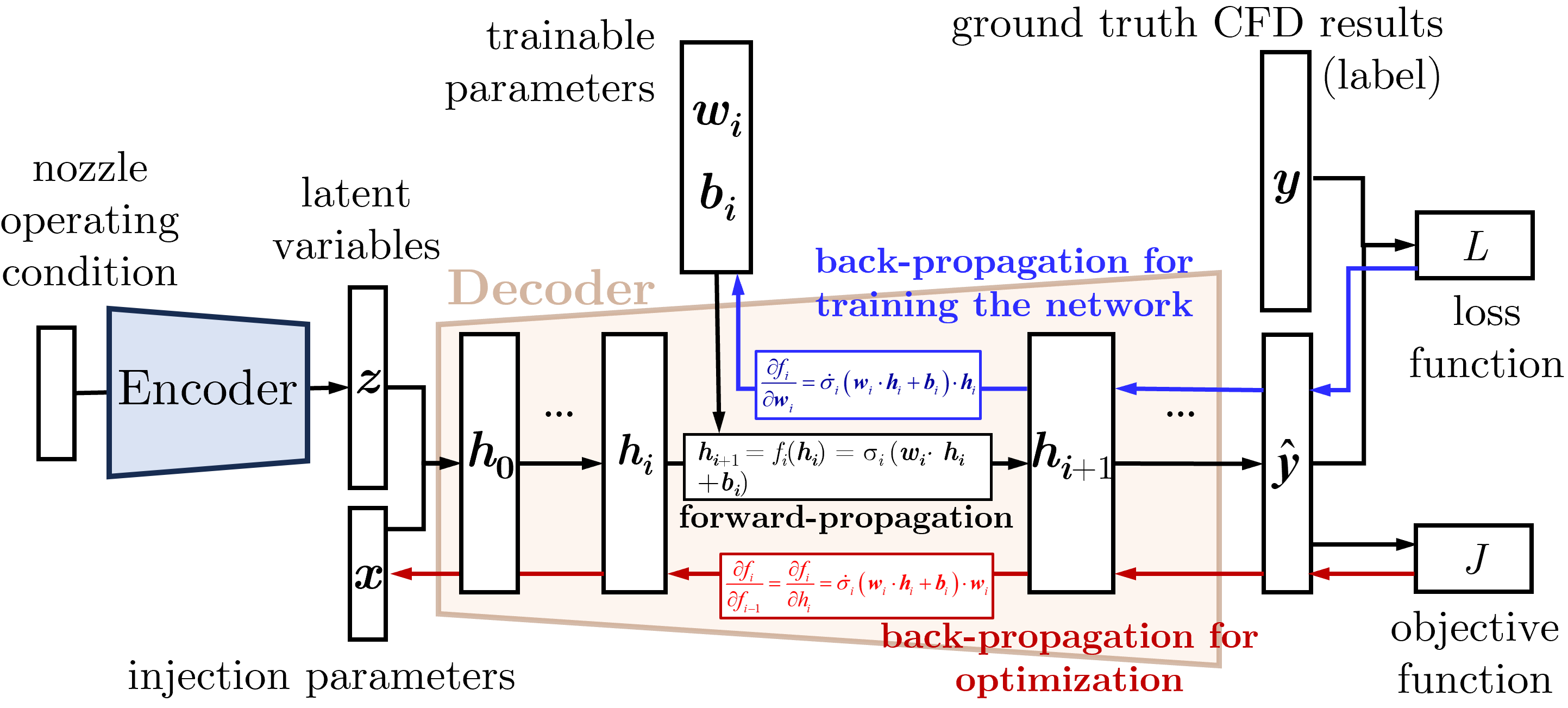}
    \caption{Data flow of forward and backpropagation in the proposed framework}
    \label{fig:bp}
\end{figure}

Figure \ref{fig:bp} shows the data flow during forward and backward propagation in the present framework. The model maps input $x$ (i.e., injection condition parameters) to output $\hat y$ (i.e., thrust coefficients). The mapping function can be represented as $\hat y = f_\theta(x)$, where $\theta$ is the trainable parameter in the network. $f_\theta(x)$ contains multiple differentiable functions:
\begin{equation}
    \hat y = f_\theta (x)=\left(f_{\theta_1}^{(1)} \circ f_{\theta_2}^{(2)} \circ \cdots \circ f_{\theta_k}^{(k)}\right)(x)
\end{equation}
where $f_{\theta_i}^{(i)}$ parameterized by weights $w_i$ and biases $b_i$. During training, the output is obtained from the inputs via forward-propagation, and the gradients of the loss function (error measure) $\mathcal L = MSE(y - \hat y)$ with respect to the parameters are computed via the chain rule:
\begin{equation}
    \frac{\partial \mathcal L}{\partial w_i} = \frac{\partial \mathcal L}{\partial \hat y}\cdot \dot f^{(k)} \cdots \dot f^{(i+1)} \cdot \frac{\partial f^{(i)}}{\partial w_i}, \quad \dot f^{(j)} = \left.\frac{\partial f^{(j)}_{\theta_{j}}}{\partial f^{(j-1)}_{\theta_{j-1}}} \right|_{\hat y^{(j-1)}(x)}
\end{equation}
where the accumulated value of the component function $\hat y^{(j-1)}(x)$
is obtained and stored during the forward-propagation process. 


Once trained, the same BP method can also be used to compute the gradient of the optimization objective $\mathcal J = \mathcal{J}(\hat y)$ with respect to the design parameters $x$:
\begin{equation}
	\frac{\partial \mathcal{J}}{\partial x}=\frac{\partial \mathcal{J}}{\partial f} \dot f^{(k)} \cdots \dot f^{(2)} \cdot \frac{\partial f^{(1)}}{\partial x}
\end{equation}

Compared with the traditional finite difference (FD) method, BP greatly reduces the need to call the model to evaluate the performance function across multiple inputs, thereby accelerating the optimization process.

\subsubsection{Gradient-based optimization framework}

\begin{figure}[ht]
    \centering
    \includegraphics[width=15cm]{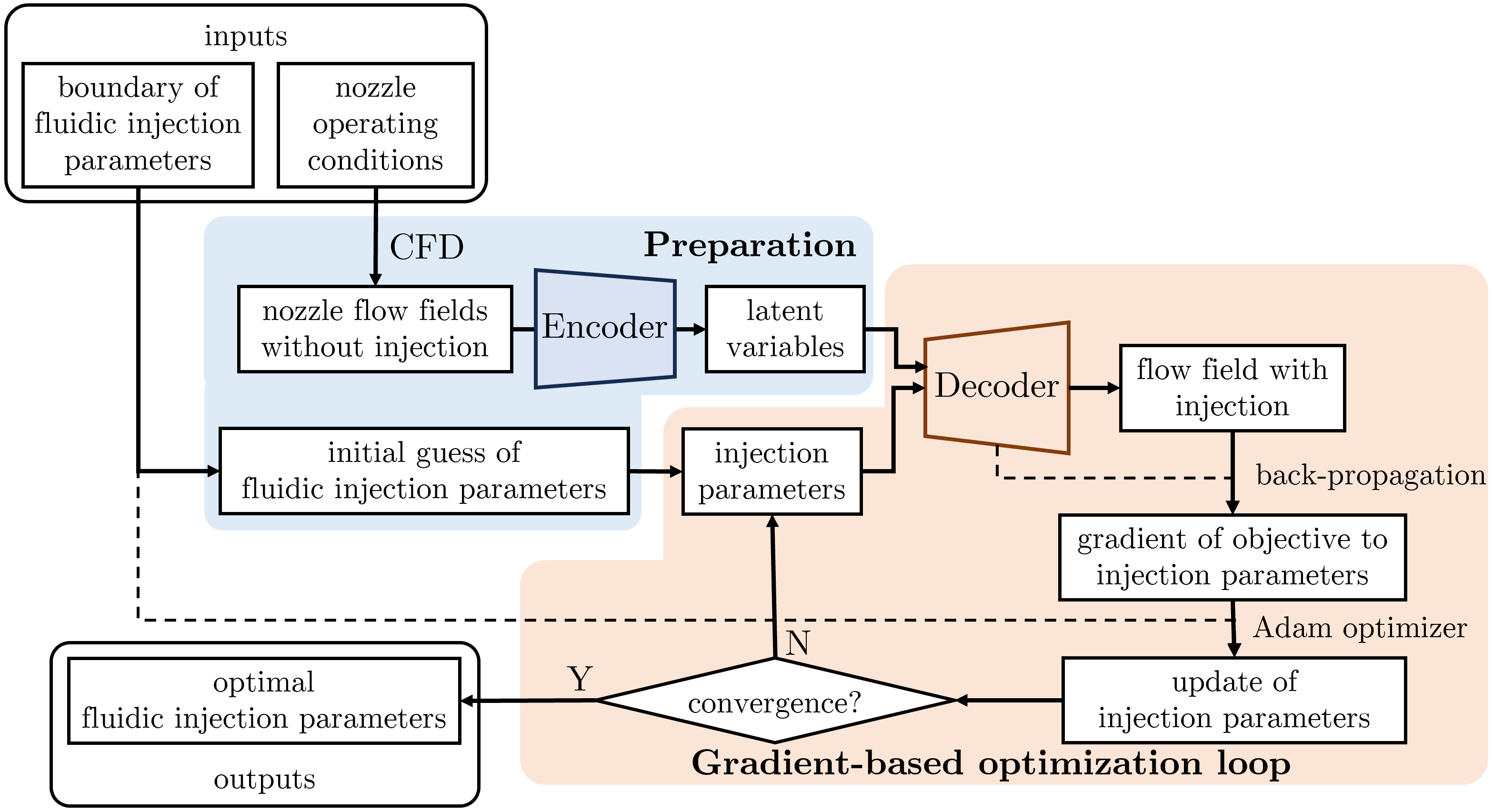}
    \caption{Gradient-based optimization procedure via back-propagation of a neural network}
    \label{fig:pro}
\end{figure}

Figure \ref{fig:pro} summarizes the gradient-based optimization procedures via BP on the machine learning model. The procedure inputs are the nozzle operating conditions $\mathcal C$ and the boundaries of the injection parameters. For each nozzle operating condition, a CFD simulation is conducted to obtain the nozzle flow fields without fluidic injection, which serve as the inputs to the pUNet model to predict with-injection flow fields. The simulation results are fed into the \textit{Encoder} and the latent variables corresponding to each nozzle operating condition are obtained and stored. 

Then, the gradient-based optimization loop starts. The trained decoder is used to predict the nozzle flow field for given injection parameters under each nozzle operating condition, and an objective function is derived as the average thrust coefficient across conditions. Next, the aforementioned BP is used to obtain the gradients. Since the constraints are simple in this problem, the Adam optimizer is used with a cosine learning rate schedule. The initial learning rate is 0.1, and the termination criterion is that the residual reaches $1 \times 10^{-5}$ or after 30 steps.

\subsubsection{Advantages}

\begin{itemize}
    \item Multipoint optimization: Compared with CFD-based optimization, whose time cost scales linearly with the design points, the proposed framework is almost independent of the problem dimension. During the preparation stage, increasing the number of conditions requires an additional CFD simulation for each condition, which can be parallelized. For the optimization loops, the prediction and back-propagation can also be performed in parallel. Owing to the underlying acceleration mechanism for matrix operations in graphics processing unit calculations, it will not impose a significant computational burden until the memory limit is reached.
    \item Multistart strategy: The high speed of machine learning-based forward and backward prediction also enables the use of the \textit{multistart strategy} to overcome the relatively poor global search capability of gradient-based methods. It is possible to use a relatively large number of starting points in the optimization, and the best results across rounds can be used as the solution to the optimization problem.
\end{itemize}

\subsection{Optimization setup}

As a test case, Table \ref{tab:testocs} provides the seven nozzle operating conditions selected through the flight envelope as the design points. The design parameters include seven $SPR$s under each nozzle operating condition and the one-value injection location, angle, and total inlet temperature. This yields a total of 10 design parameters. Their design space is the same as the range of the training database, as shown in Table \ref{tab:ranges}. In this test case, the plain average thrust coefficient among the seven nozzle operating conditions is recognized as the optimization objective.

\begin{table}[ht] 
\small 
\centering
\caption{Nozzle operating conditions at the seven design points.}\label{tab:testocs}
\begin{tabular}{cccccccc}
\toprule
Design point index & 1 & 2 & 3 & 4 & 5 & 6 & 7\\
\midrule
$Ma$ & 1.5 & 1.7 & 1.9 & 2.1 & 2.3 & 2.5 & 2.7\\
$H$ (km) & 15.0 & 15.5 & 16.0 & 16.5 & 17.0 & 17.5 & 18.0\\
$NPR$ & 5.0 & 7.5 & 10.0 & 12.5 & 15.0 & 17.5 & 20.0\\
$T_7^*$  (K) & 1367 & 1380 & 1395 & 1411 & 1429 & 1448 & 1467\\
\bottomrule
\end{tabular}
\end{table}

20 initial guesses of the injection parameters are sampled along the boundary via Latin hypercube sampling (LHS). They serve as the starting points of the optimization. During optimization, a small initial injection intensity leads to a local optimum in which the injections are eliminated. Therefore, the initial values of the SPRs are limited to greater than 0.5.

\subsection{Optimization results}

\subsubsection{Overall results and optimization process}

\begin{figure}[ht]
    \centering
    \includegraphics[width=0.5\linewidth]{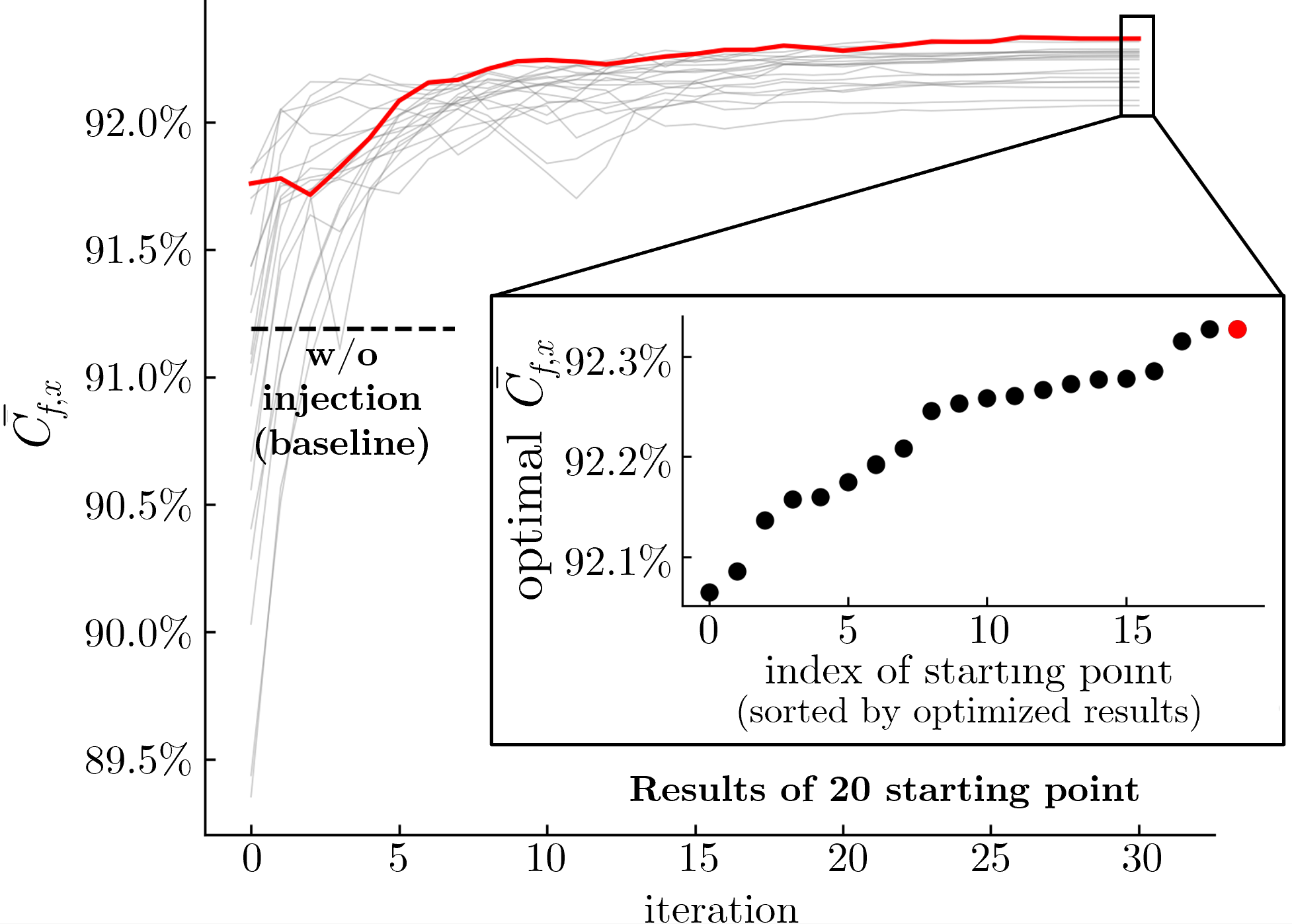}
    \caption{Optimization process and results of the 20 starting points}
    \label{fig:opt}
\end{figure}

Figure \ref{fig:opt} shows the optimization processes and results of the average thrust coefficient for the 20 optimization starting points. The round with the largest objective result is shown in red in both figures. The fluidic injection with the best-optimized parameters improves the average thrust coefficient by 1.14\% relative to non-injection conditions. Since the non-injection thrust coefficient is close to 1 for operating conditions with a large $NPR$, achieving such optimization results in this optimization case is good. The results also align with the conclusions in other papers that investigated the influence of injection parameters

The results of the 20 rounds are scattered within a range of 0.26\%, which is relatively large given the improvement. The results also show a cluster pattern, suggesting several local optima, each corresponding to a cluster. This justifies the multistart strategy for gradient-based optimization.

\begin{figure}[H]
    \centering
    \includegraphics[width=0.8\linewidth]{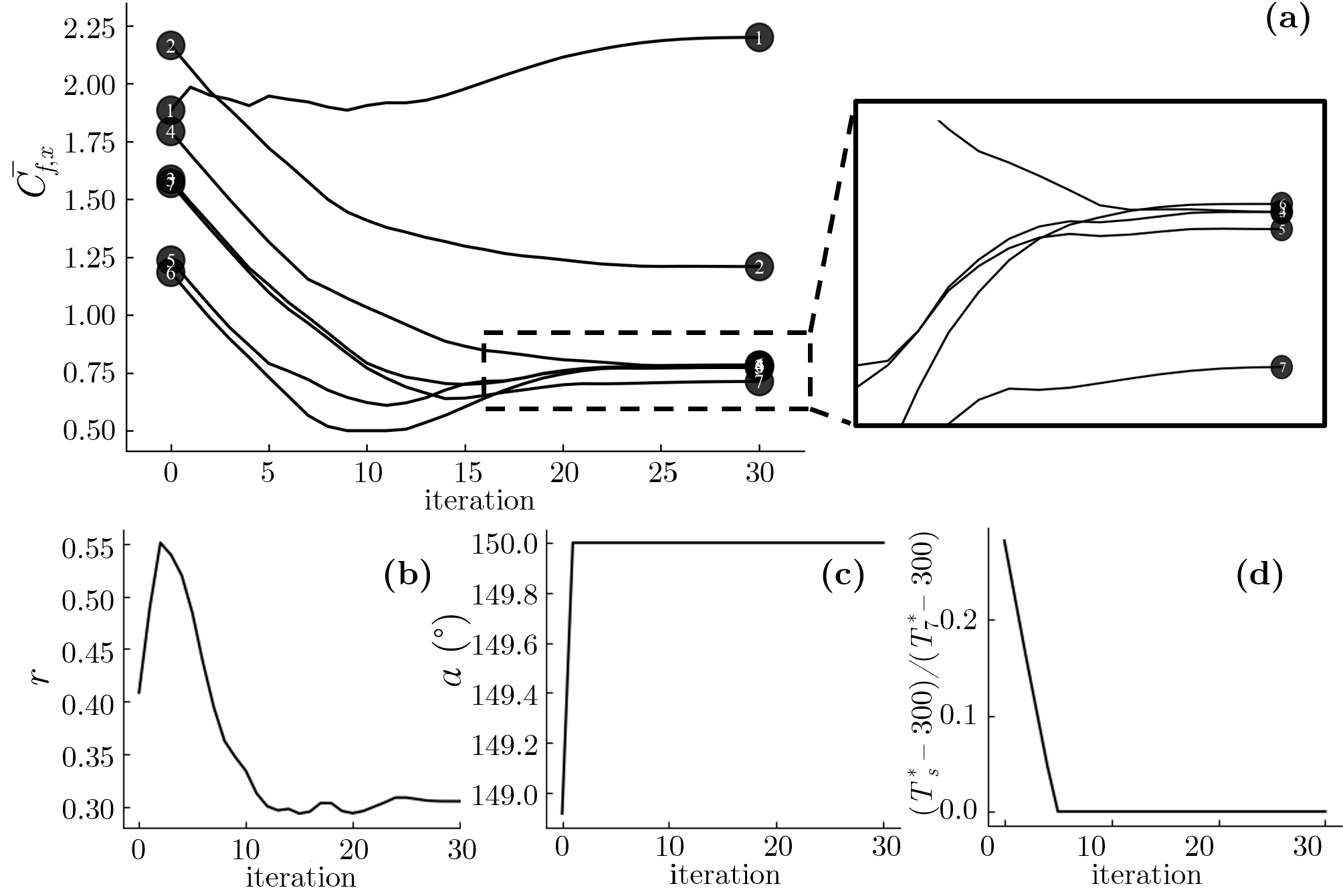}
    \caption{Injection parameters during the best optimization round (\textbf{a}) Pressure ratios of 7 nozzle operating conditions. (\textbf{b}) Injection position. (\textbf{c}) Injection angle. (\textbf{d}) Injection temperature ratio.}
    \label{fig:paras}
\end{figure}

Figure \ref{fig:paras} shows the evolution of the design variables during the optimization process. The optimal injection intensity, $SPR$, decreases as the $NPR$ increases, and the values for the last five conditions converge to nearly the same level. The injection location moves upstream, while the injection angle reaches its maximum, and the temperature decreases to its minimum. This trend is consistent with engineering intuition, as all changes improve injection efficiency for a given injection intensity. Since the injection intensity is included in the calculation of the total ideal thrust when evaluating the thrust coefficient, it is reasonable for the optimization to maximize injection efficiency. 

\subsubsection{Verification of the best injection parameters}

\begin{figure}[ht]
    \centering
    \includegraphics[width=0.55\linewidth]{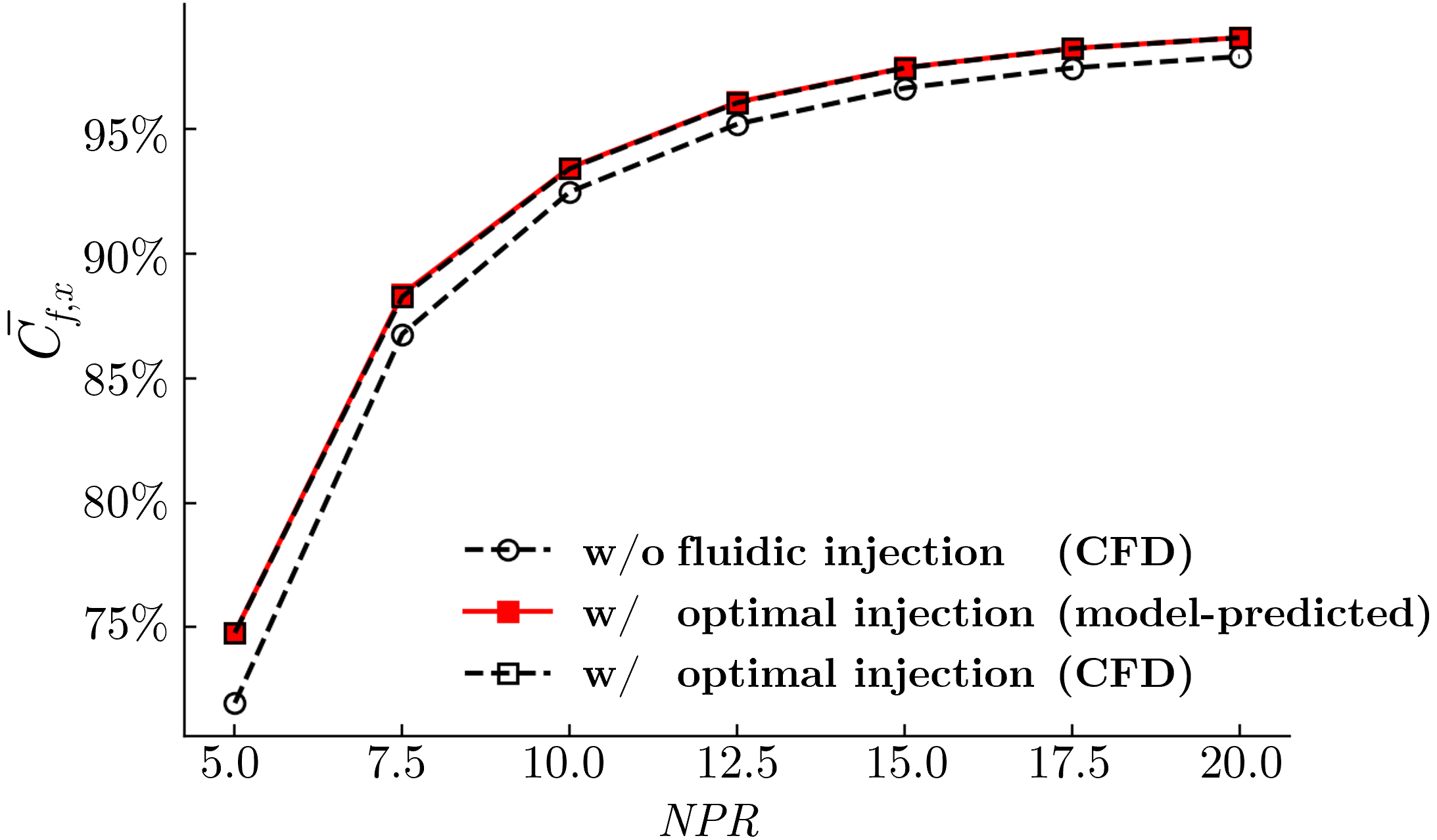}
    \caption{CFD verification of the optimal injection performance}
    \label{fig:vali}
\end{figure}

The performance improvements by the best injection parameters are verified with CFD simulations, and the thrust coefficients are shown against the $NPR$s in Fig. \ref{fig:vali}. The CFD-simulated and model-predicted thrust coefficients for optimal injection are very close, with an average prediction error of around 0.0003, which is on the same scale as the test database. 

\begin{figure}[h]
    \centering
    \includegraphics[width=\linewidth]{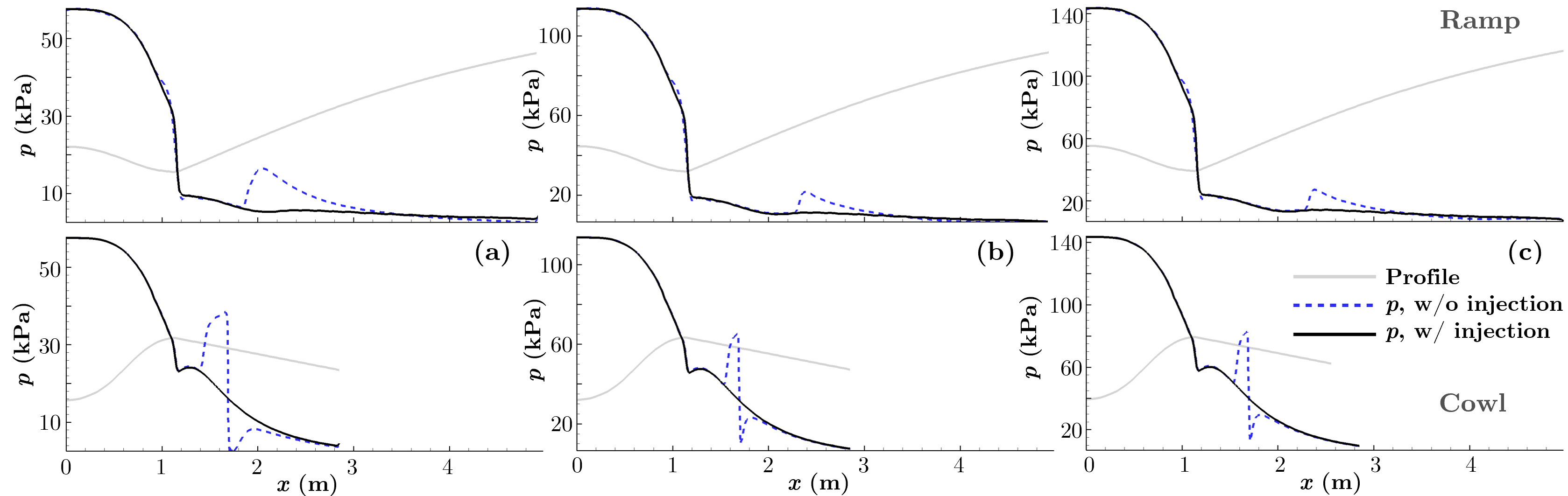}
    \caption{Pressure distributions on the nozzle surfaces with and without optimized injection (\textbf{a}) Design point 1 ($NPR = 5.0$). (\textbf{b}) Design point 4 ($NPR = 12.5$). (\textbf{c})  Design point 7 ($NPR = 20.0$).}
    \label{fig:dist}
\end{figure}

Figure
\ref{fig:dist} visualizes the optimized fluidic injection at three different design points. The shock waves induced by the injection form high-pressure regions on both the ramp and cowl surfaces, which contribute to a greater force in the x-direction. The optimization found a joint optimal solution for the injection intensities. The overexpansion is severe for design points with low $NPR$ (in subfigures (a)), so stronger injections are adopted, which create stronger shock waves. The injection intensities are small for design points with larger $NPR$s, since stronger injections yield only a marginal advantage.


\subsubsection{Comparison between BP and AD}

The optimization results are also compared with those of the traditional FD method to verify the gradients obtained with the BP method. The same optimization problem setup and initial points are used, and the gradients used for this case are obtained via the FD method from the same pretrained model. The FD method is implemented with the open-source library \texttt{SciPy}, where the epsilon for the finite differential is $1 \times 10^{-5}$.

The optimization results with the BP gradients (labeled ML+BP) and with the FD gradients (labeled ML+FD) are compared in Table \ref{tab:fd}. The two optimizations achieve the best result in the round that starts from the same initial points among the 20 multistart runs, and the improvements are almost identical. The proposed BP-based method even achieves an additional improvement of 0.005\%, attributed to numerical errors in the FD method. The average improvement of the BP-based method among the 20 multistart points also exceeds that of the FD-based method. These results justify the BP gradients in optimization.

\begin{table}[ht] 
\small 
\caption{Comparison of optimization results using BP and FD gradients.}\label{tab:fd}
\centering
\begin{tabular}{ccc}
\toprule
 & ML+BP & ML+FD \\
\midrule
Best optimized $\bar C_{f,x}$ (among 20 starting points) & +1.140\% & +1.135\% \\
Average optimized $\bar C_{f,x}$ (among 20 starting points) & +1.104\% & +1.103\%\\
\bottomrule
\end{tabular}
\end{table}

\subsubsection{Comparison to single point optimization}


The single-condition optimization only considers the thrust coefficient under the first nozzle operating condition, i.e., $H=$ 15 km, $Ma=$ 1.5, $NPR=$ 5.0, and $T_7^* = $ 1367 K. The optimization yields the optimal injection parameters: a secondary pressure ratio of 2.375, an injection position ratio of 0.126, an injection angle of 150°, and an injection temperature ratio of 0.0. This set of parameters is used for the other nozzle operating conditions.

\begin{figure}[ht]
    \centering
    \includegraphics[width=0.5\linewidth]{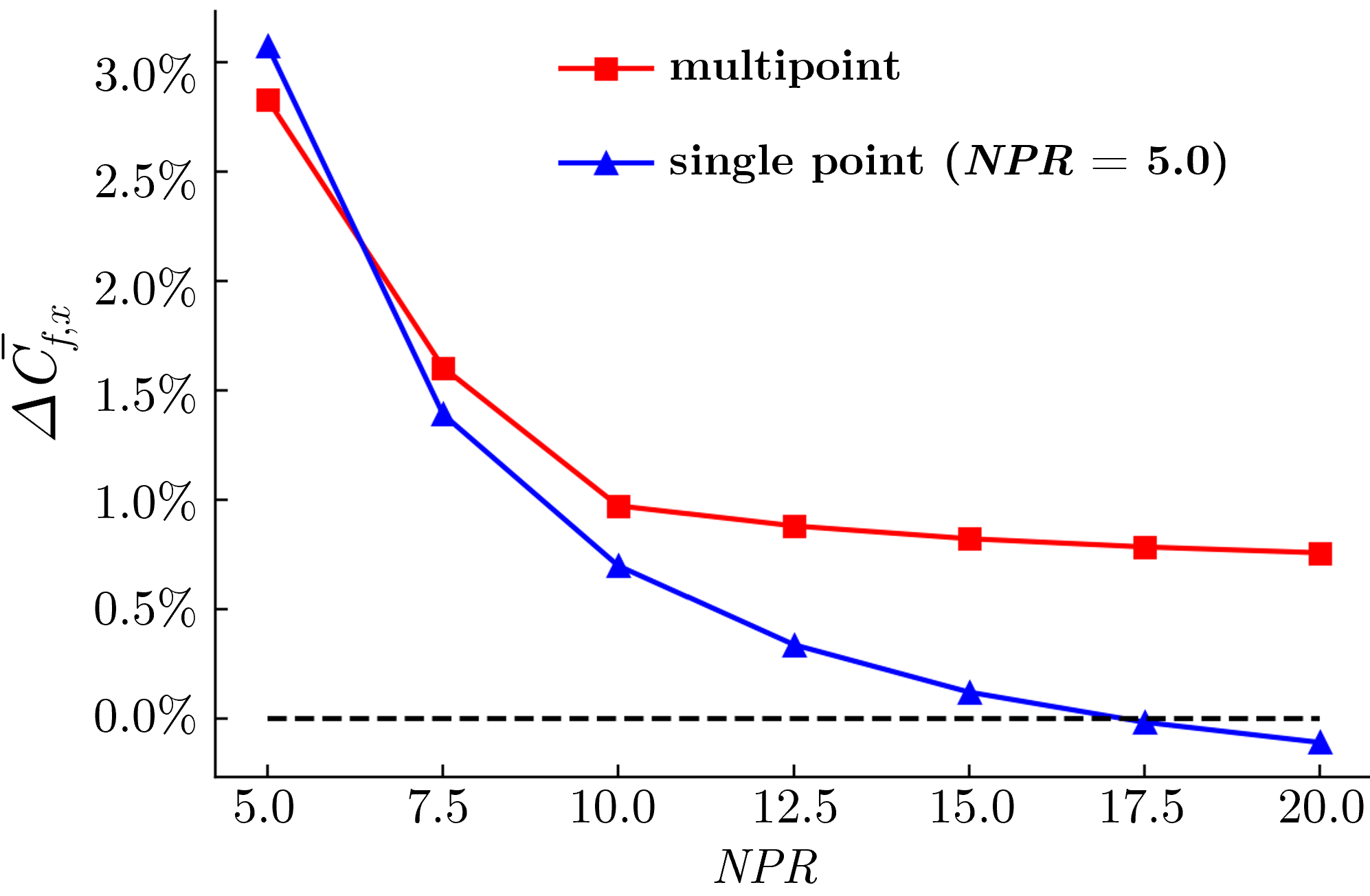}
    \caption{Comparison of single-point and multipoint optimization results}
    \label{fig:somo}
\end{figure}

The overall thrust improvements for single-point and multipoint optimization are shown in Fig. \ref{fig:somo} with respect to $NPR$. The thrust improvement from single-point optimization exceeds that of multipoint optimization only at the selected point, whereas it is lower than that of multipoint optimization under other nozzle operating conditions. The injection may even negatively affect performance when $NPR$ exceeds 17.5.

\subsubsection{Comparison of computational costs}

\begin{figure}[ht]
    \centering
    \includegraphics[width=0.6\linewidth]{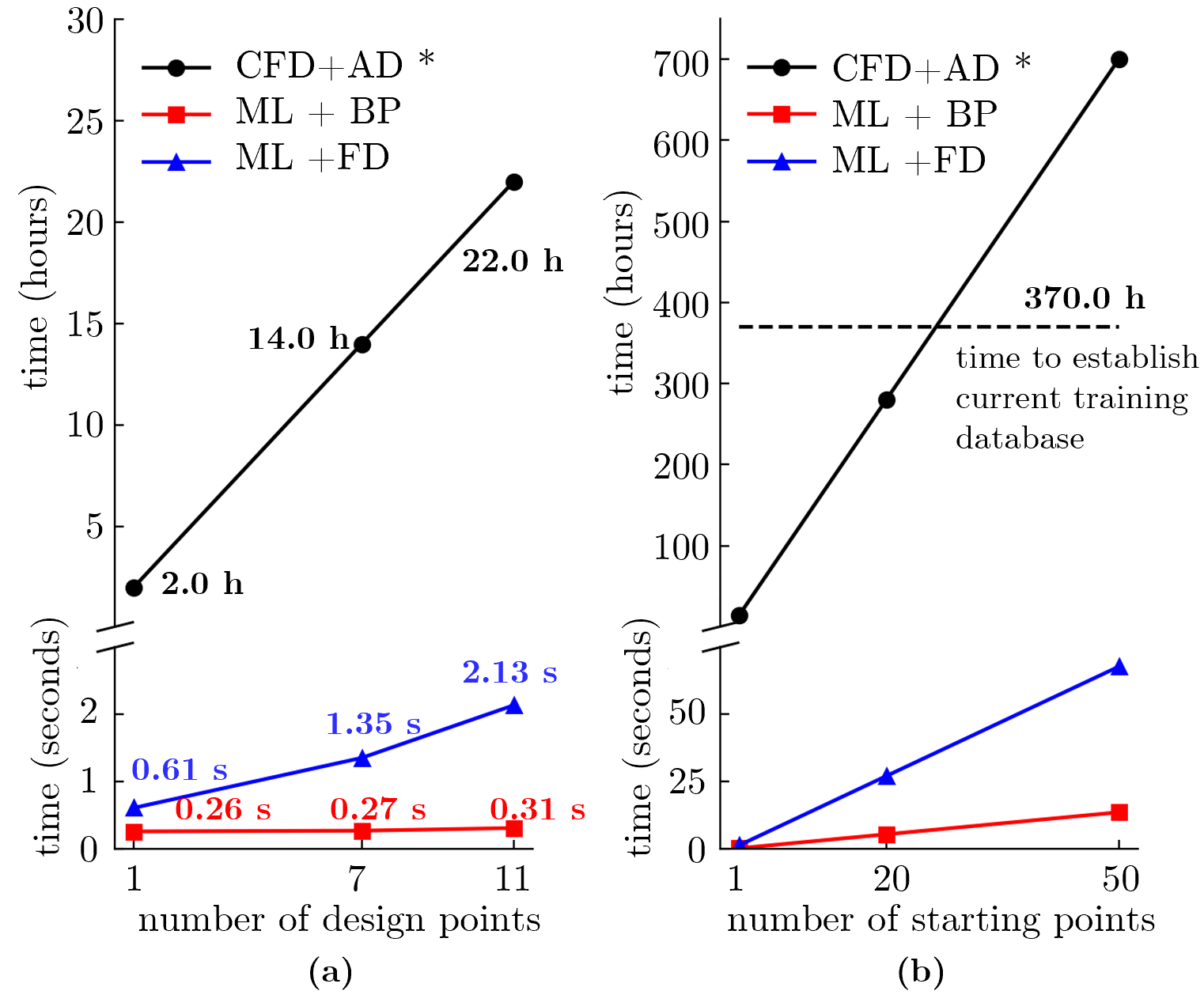}
    \caption{Time cost evaluation for optimization based on different methods (\textbf{a}) Computation cost for a single round with respect to the number of design points. (\textbf{b}) Computation cost for 7-design-point optimization with respect to the number of starting points. * The CFD + AD time cost is estimated on the basis of the similar time taken for CFD simulation and adjoint solution.}
    \label{fig:time}
\end{figure}

Here, the key advantage of the proposed methodology, the computational cost, is compared with traditional methods. In the above section, the multipoint evaluation and multistart strategies are both demonstrated to be necessary, and the associated time costs are analyzed. Figure \ref{fig:time} shows the computational costs for optimization with different numbers of design points and different numbers of starting points. The time cost for the proposed ML + BP is approximately 0.2-0.3 seconds and remains nearly constant as the number of design points increases. For comparison, with the traditional AD method for gradient computation from a pre-trained model, the time cost per forward-propagation step remains almost the same. As the number of design points increases, the number of model forward calls increases linearly. The time cost of the ML + FD method for one round of optimization thereby increases nearly linearly.

The CFD-based optimization relies on the adjoint method to compute the gradient, requiring one CFD simulation and one adjoint computation for each group of injection parameters. Given that one CFD simulation takes approximately 2 minutes and that solving for the adjoint problem takes a similar amount of time, the estimated time cost of single-point optimization is approximately 2.0 CPU hours for one starting point. However, as the number of design points or multistart points increases, the time cost scales linearly, resulting in a high cost for multistart multipoint optimization even when CFD simulations can be conducted in parallel. For optimization with seven design points and 20 starting points, the time cost approaches that for establishing the current training database. Given that the time invested in building the database can be shared across different optimization problems, the proposed framework is even more efficient than the CFD-based one.

\section{Conclusion}

Using a machine learning model and a back-propagation algorithm, the fluidic injection parameters are optimized for the first time under multiple nozzle operating conditions. The proposed methodology is demonstrated to be effective and efficient, and can be easily applied to other optimization settings as needed. The key contributions of this study can be summarized as follows:

\begin{enumerate}
    \item Given that fluidic injection affects only the flow field in the vicinity of and downstream of the injection slot, the prior-based prediction strategy is applied. A U-Net is designed to predict the differences in distributions with and without injection. On the test dataset, the model precisely predicts surface distributions and achieves an average prediction error of 0.03\% for the thrust coefficient, making it reliable for optimization.
    \item An efficient optimization framework is built for multipoint injection parameter optimization. The pre-trained model is used to evaluate the objective function (the average thrust coefficient) and the gradients via the back-propagation algorithm. With 20 starting points, the optimization increased the average thrust coefficient by 1.14\%, as verified by CFD simulations. 
    \item Further studies justify the need to conduct multiple-design-point optimization to achieve better overall performance and to use the multistart strategy to avoid local optima. Compared with traditional methods, the proposed data-driven method has a computational cost that is almost independent of the number of design points, thereby greatly reducing the time cost of multipoint optimization with a multistart strategy, even when the cost of establishing the database is considered.
\end{enumerate}

Overall, this paper presents a practical methodology for multipoint optimization problems and illuminates potential applications of machine learning techniques in the design of aerospace propulsion systems.

\section{Data Availability Statement}

The dataset to train the model is available at \href{https://huggingface.co/datasets/yunplus/nozzleinjection}{https://huggingface.co/datasets/yunplus/nozzleinjection}. The source code can be found at \href{https://github.com/YangYunjia/floGen}{https://github.com/YangYunjia/floGen} under \texttt{example/nozzle}.

\section{Acknowledgments}

This work was supported by the 1912 project and National Natural Science Foundation of China Nos. 92052203, 12372288, and 12388101. The authors would like to thank Runze Li for the helpful discussions.

\section{Conflicts of interest}

The authors declare no conflicts of interest.

\section{Notation}
\label{app:notation}
\emph{The following symbols are used in this paper:}
\nopagebreak
\par
\begin{tabular}{r  @{\hspace{1em}=\hspace{1em}}  l}
$A$         & Area (height, in 2D cases); \\
$C_{f,x}$   & $x$-direction thrust coefficient; \\
$F$         & Thrust; \\
$F_{\textrm{id}}$ & Ideal thrust; \\
$H$         & Flight height; \\
$p$         & Static pressure; \\
$p_{\textrm{atm}}$  & Ambient pressure; \\
$r$         & Ratio of the injection slot on the cowl surface; \\
$\bm V$     & Velocity; \\
$\alpha$    & Injection angle; \\
$\gamma$    & Heat ratio; \\
$\bm \Phi$  & Momentum flux; \\
$Ma$        & Mach number; \\
$NPR$       & Nozzle pressure ratio ($=p_7^*/p_{\mathrm{atm}}$; \\
$SPR$       & Secondary pressure ratio ($=p_s^*/p_7^*$); \\
$(\cdot)'$  & Values after injection; \\
$(\cdot)^*$ & Total value; \\
$(\cdot)_{7,8,9}$  & Values at inlet, throat, and exit plane, respectively;\\
$(\cdot)_{s}$  & Values at injection;\\
$\tilde{(\cdot)}$ & Nondimensional values.

\end{tabular}

\pagebreak
%
\appendix
%
%

\section{Calculation of the thrust coefficient from the predicted distributions}\label{app:coef}

There are several extra steps to calculate the thrust coefficient $C_{f,x}$ from the predicted distributions. The nozzle thrust after injection for a discrete surface pressure distribution can be derived from Eq. \ref{eqn:injf} as
\begin{equation}
    \bm{F}_{\mathrm{inj}} \approx \bm {\Phi}_8' + \sum_{i=1}^{N_{\mathrm{point}}} (p'_i - p_{\mathrm{atm}}) \bm n_i l_i + \bm {\Phi}_s,
\end{equation}
where $n_i$ and $l_i$ are the unit normal vectors and the length of each discrete segment, which are precalculated and stored, and $p_i'$ is the average pressure. For simplicity, the throat momentum flux in the noninjection case is used, i.e., $\bm {\Phi}_8' = \bm {\Phi}_8$. It can be calculated by integrating the throat pressure and velocity distributions. The momentum flux through the injection slot ($\bm \Phi_{\mathrm{inj}}$) is obtained with an isentropic relationship. Its direction is the same as the injection angle, and its magnitude is
\begin{align}
    ||\bm {\Phi}_s||&= m_s V_s\\
    m_s &= \sqrt{\frac{\gamma}{R}\left(1+\frac{\gamma-1}{2}Ma_s^2\right)^{-\frac{\gamma+1}{\gamma-1}}}\cdot Ma_s \cdot \frac{p_s^*}{\sqrt{T_s^*}}A_s\\ V_s &= Ma_s \cdot \sqrt{\gamma R\cdot \frac{T_s^*}{1 + \frac{\gamma-1}{2}}Ma_s^2}
\end{align}
where $Ma_s = 1.0$ is the Mach number at the exit plane of the injection slot, given the throat property. $A_s = w \cdot \sin \alpha$ is the exit plane area, which is the projected length of the slot width. In Equations, $p_s^*$ and $T_s^*$ are the total inlet conditions of the injection. 

Then, the thrust coefficient is derived with $C_{f,x}' = F_{x}' / (F_{\mathrm{id}} + F_{\mathrm{id}}')$.

%
%
\bibliography{ascexmpl-new}

\end{document}